\documentclass[aip,preprint]{revtex4-1}

\usepackage{graphicx}
\usepackage{dcolumn}

\usepackage{amssymb}
\usepackage{amsmath}
\usepackage{color}

\draft 

\begin{document}

\title{An $O(N)$
and parallel approach to integral problems by a kernel-independent fast multipole method: Application to polarization and magnetization of interacting particles}

\author{Xikai Jiang}
\email[Email: ]{xikai@anl.gov}
\affiliation{Materials Science Division, Argonne National Laboratory, Lemont, Illinois 60439, USA}

\author{Jiyuan Li}
\affiliation{Institute for Molecular Engineering, University of Chicago, Chicago, Illinois 60637, USA}

\author{Xujun Zhao}
\affiliation{Mathematics and Computer Science Division, Argonne National Laboratory, Lemont, Illinois 60439, USA}

\author{Jian Qin}
\affiliation{Department of Chemical Engineering, Stanford University, Stanford, California 94305, USA}

\author{Dmitry Karpeev}
\email[Present address: ]{KCG Holdings, Inc.}
\affiliation{Mathematics and Computer Science Division, Argonne National Laboratory, Lemont, Illinois 60439, USA}
\affiliation{Computation Institute, University of Chicago, Chicago, Illinois 60637, USA}

\author{Juan Hernandez-Ortiz}
\affiliation{Institute for Molecular Engineering, University of Chicago, Chicago, Illinois 60637, USA}
\affiliation{Department of Materials and Minerals, Universidad Nacional de Colombia--Medellin, Medellin, Colombia}

\author{Juan de Pablo}
\affiliation{Materials Science Division, Argonne National Laboratory, Lemont, Illinois 60439, USA}
\affiliation{Institute for Molecular Engineering, University of Chicago, Chicago, Illinois 60637, USA}

\author{Olle Heinonen}
\email[Email: ]{heinonen@anl.gov}
\affiliation{Materials Science Division, Argonne National Laboratory, Lemont, Illinois 60439, USA}
\affiliation{Northwestern-Argonne Institute for Science and Engineering, Evanston, Illinois 60208, USA}

\date{\today}

\begin{abstract}
Large classes of materials systems in physics and engineering are governed by magnetic and electrostatic interactions. Continuum or mesoscale descriptions of such systems can be cast in terms of integral equations, whose direct computational evaluation requires
$O(N^{2})$ operations, where $N$ is the number of unknowns. Such a scaling, which arises from the many-body nature of the relevant Green's function, has precluded wide-spread adoption of
integral methods for solution of large-scale scientific and engineering problems. In this work, a parallel computational approach is presented that relies on using scalable open source libraries and utilizes a kernel-independent Fast Multipole
Method {(FMM)} to evaluate the integrals in $O(N)$ operations, with $O(N)$ memory cost, thereby substantially improving the scalability and efficiency of computational integral methods. We demonstrate
the accuracy, efficiency, and scalability of our approach in the contest of two examples. In the first, we solve a boundary value problem for a
ferroelectric/ferromagnetic volume in free space.
In the second, we solve an
electrostatic problem involving polarizable dielectric bodies in
an unbounded dielectric medium. 
The results from these test cases show that our proposed parallel approach, which is built on a kernel-independent FMM, can enable highly efficient and accurate simulations and allow for considerable flexibility in a broad range of applications.
\end{abstract}

\pacs{Valid PACS appear here}


\maketitle 

\section{Introduction}

Massively parallel computer hardware, and the corresponding software, have enabled solution of increasingly complex problems in science and engineering. Such problems are often cast in terms of partial differential equations (PDEs). In many cases, it is convenient to formulate the PDEs in terms of integral equations, which can then be solved by relying on a variety of numerical methods. For the particular case of boundary integral equations, the Boundary Element Method (BEM)\cite{bem1984} has found numerous applications in physical problems ranging from solids and fluids\cite{bie_solid_fluid},
to multiphase materials\cite{bie_multiphase2001}, heat transfer\cite{heat_transfer2000}, electrostatics\cite{allen2001}, and magnetostatics.
For example, BEMs were used in fluid problems to study the motion and deformation of bubbles\cite{Zhang2014} and 
to determine the phase diagram of complex fluids\cite{phase2005};
BEMs have also been used to address three-dimensional (3D) linear elasticity problems involving particles embedded in a binder \cite{gmres_fmm_1998}, and in crack analysis in 3D time harmonic elastodynamics\cite{crack1989}.
BEMs are particularly well-suited for problems that involve bodies embedded in an infinite medium, with the boundary condition that the relevant potential decays to zero at infinity. If the embedding space is homogeneous and the governing equations for the underlying physics are linear,
BEMs are frequently used in order to avoid explicitly including the embedding space exterior to the bodies.
In this work, we consider two versions of this kind of problem as explicit test cases: interacting ferromagnetic bodies embedded in space, and dielectric bodies embedded in an infinite dielectric medium. However, our proposed method is widely applicable to problems in which the mathematical formulation can be cast in terms of a boundary integral equation.

For the problem of interacting ferromagnetic bodies embedded in space, the interactions between
the bodies are calculated using an open boundary condition on their boundaries. Specifically, the non-zero, finite magnetization of the bodies leads to a discontinuity in the magnetostatic potential at their boundaries.
One approach to this problem is to use a hybrid finite element (FEM)--BEM
to calculate the non-local magnetostatic interactions between the magnetic bodies\cite{fk1990}.
The method has the advantage that it allows for arbitrary-shaped bodies and requires no mesh between them. A disadvantage is that the
integral of the Green's function over the boundary of all magnetic bodies leads to a
computational complexity of $O(N^2)$ in direct implementations, where $N$ is the number of
degrees of freedom (DoFs) on the boundaries of all magnetic bodies.
Several techniques have been proposed to increase the efficiency of the boundary integral calculation.
Examples include the tree-code algorithm and the hierarchical matrices method\cite{micromag2003, micromag2009}, which exhibit computational complexities of $O(N\log N)$ and $O(N)$, respectively.

In electrostatics, a similar class of problems is concerned with the electrostatic polarization of finite-size dielectric bodies
embedded in an infinite dielectric medium. Free charges carried by these bodies, or free point charges in the medium,
are sources of electrostatic fields. Because the relative permittivity of the medium is usually different from those
of the bodies, the electrostatic field is discontinuous across the interface between a body and the medium.
As a consequence, bound surface charges are induced on the interface
to compensate for the discountinuity in the electrostatic field. In order to calculate the induced bound charges
on the interface, a BEM was originally proposed using a
variational approach\cite{allen2001} to solve the underlying Poisson equation. In numerical simulations of the polarizable bodies, BEMs have been used for stationary or mobile dielectrics\cite{boda2004,2010CHolm,2012_dirk,luijten2014,luijten2014jcp}.
A central aspect of this class of problems is the calculation of the distribution of bound surface charge density through solutions of a
linear system of equations $Ax=b$, where $A$ is a dense matrix whose entries depend on the geometry of
the dielectric bodies, $x$ represents the induced bound surface charge density, and $b$ represents contributions from free charges
on the dielectric bodies, or free point charges in the exterior medium.
The iterative Generalized Minimal Residual Method (GMRES)\cite{gmres} provides an efficient numerical method to
solve the resulting dense linear system of equations\cite{luijten2015jcp}. However, the matrix-vector multiplication
in each GMRES iteration has a computational complexity of $O(N^2)$ in any direct implementation. In Ref.~
\onlinecite{luijten2014jcp,luijten2015jcp}, the matrix-vector multiplication is accelerated using a fast Ewald solver,
i.e., the particle-particle particle-mesh method (PPPM or P$^3$M)\cite{p3m1973}, which offers a computational complexity of $O(NlogN)$.

In this work, we have
developed an efficient and scalable parallel computational approach for BEMs that is built on the scalable parallel libraries libMesh (http://libmesh.github.io)\cite{libmesh2006} and ScalFMM (http://scalfmm-public.gforge.inria.fr)\cite{scalfmm}.
The boundary integrals are 
accelerated using a kernel-independent fast multipole method (FMM)\cite{Fong2009} with $O(N)$ computational
complexity. This kernel-independent FMM is an interpolation-based FMM that utilizes Chebyshev interpolation
and low-rank approximation, and can also be used for kernels that are known only numerically.
It was first developed for non-oscillatory kernels and later extended for oscillatory
kernels\cite{Messner2012}.
Our approach to boundary integrals also adopts a matrix-free method that requires no explicit storage of
a global matrix, thus enabling simulations with a memory cost of only $O(N)$. The corresponding software is
is made freely available to potential users.
In the remainder of this paper,
we first introduce the general boundary integral problem. We then discuss methods for the magnetostatic and electrostatic problems, and our implementations of BEM;
We apply our computational approach to several model problems and test its accuracy, efficiency, and scalability. We conclude with a summary and suggestions for future directions for improvements
of the proposed approaches, and elaborate on several research areas and physical phenomena to which our approach could be applied. 

\section{Methods}

\subsection{General integral problems}

\begin{figure}[h]
\includegraphics[width=0.35\textwidth]{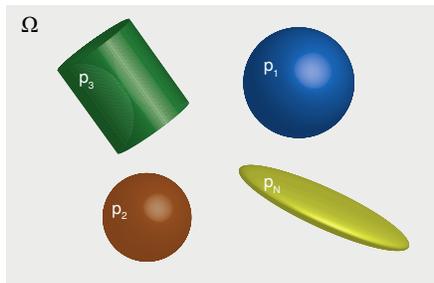}
\centering
\caption{A schematic of general integral problems. $\Omega$ is the exterior region and the interior region $\Gamma$ consists of different arbitrary-shaped bodies. $\Gamma=\bigcup\limits_{i=1}^{P}p_{i}$, where $P$ is the total number of bodies and $p_{i}$ is the region occupied by $i$th body. The boundary the interior region is denoted as $\partial\Gamma$. \textcolor{red}{Confined geometries can also be considered as shown in Fig.~\ref{fig_color_dielectric}(b).}}
\label{fig4_fk}
\end{figure}

Many integral problems share similar configurations to those shown schematically in
Fig. \ref{fig4_fk}. Space is divided into an exterior region $\Omega$ and an interior region $\Gamma$ consisting of the union of different bodies with arbitrary shapes, for example dielectric or ferroelectric/ferromagnetic bodies. It is assumed here that there are no intersections or contacts between
any two bodies. In most problems, the solution is obtained from functions $f({\textbf r})$ obtained from the integral over the boundary of all bodies ($\partial\Gamma$) of a kernel $K(\textbf{r},\textbf{r}')$ and a source function $g({\textbf r})$. The kernel is derived from a Green's function that represents the underlying physics of a delta-function source; thus
\begin{equation}
f(\textbf{r})=\int_{\partial\Gamma}K(\textbf{r},\textbf{r}')g(\textbf{r}')\,d^2r'.
\end{equation}

In numerical integration of the above integral using a Gaussian quadrature rule, $f(\textbf{r}_i)$ can then in general be expressed as
\begin{equation} \label{sum_fmm}
f(\textbf{r}_i)=\sum_{j=1}^{N}K(\textbf{r}_i,\textbf{r}_j)\sigma_{j}.
\end{equation}
Equation~(\ref{sum_fmm}) is exactly the type of summation for which the FMM is used, with $O(N)$ computational complexity rather than
the $O(N^2)$ complexity that results from using the direct method over all points ${\textbf r}_i$ and ${\textbf r}_j$.
The kernel-independent FMM enables fast computation of the integral with arbitrary kernels for a variety of problems; two specific test cases are considered in the following sections.

\subsection{Magnetostatic problem}

The magnetostatic problem is ubiquitous for any system that involves patterned micron- or sub-micron size magnetic bodies, such as read or write heads in magnetic hard disk drives, magnetic nanoparticles\cite{read-write_2009}, coupled magnetic disks\cite{disks_PRB-2009,meron_prl-2012}, or artificial spin ices\cite{spin_ice-2011}. In patterned ferromagnetic systems, there typically arises a competition between the short-range exchange interactions and long-range magnetostatic interactions. These latter originate in the volume pole density $\nabla\cdot{\mathbf{M}}({\mathbf r})$ and the surface pole density $\hat n\cdot{\mathbf M}({\mathbf r})$. The magnetostatic
fields from the surface pole density then involves solving a Poisson equation with
sources on the boundary of a finite body and with the boundary condition that the
magnetic scalar potential goes to zero far away from the body.
In the framework of a hybrid FEM--BEM, the magnetic scalar potential $\phi$
is split into $\phi=\phi_{1}+\phi_{2}$, where $\phi_{1}$ solves the Poisson equation in the magnetic region (see Fig.~\ref{fig4_fk}) $\Gamma$:
\begin{equation} \label{poisson}
\nabla^{2}\phi_{1}=4\pi\nabla\cdot\textbf{M}({\textbf r}),
\end{equation}
with the boundary condition
\begin{equation} \label{bc_poisson}
\frac{\partial\phi_{1}}{\partial\textbf{n}}=4\pi\textbf{n}\cdot\textbf{M}({\textbf r}).
\end{equation}
Here, $\textbf{M}({\textbf r})$ is the magnetization density of the magnetized bodies $p_i$, and $\textbf{n}$ is the outward-pointing (from the interior of the magnetic region) unit normal vector on the boundary
of the magnetic region ($\partial\Gamma$) .
After $\phi_{1}$ is obtained by solving the Poisson equation Eq.~(\ref{poisson}),
$\phi_{2}$ is calculated first on $\partial\Gamma$ by the standard double layer integral
\begin{eqnarray} \label{fk_integral}
\phi_{2}(\textbf{r})=\frac{1}{4\pi}\int_{\partial\Gamma}\frac{(\textbf{r}-\textbf{r}')\cdot\textbf{n}(\textbf{r}')}{{|\textbf{r}-\textbf{r}'|}^3}\phi_{1}(\textbf{r}')\,d^2r'\nonumber\\
+\left(\frac{\Omega(\textbf{r})}{4\pi}-1\right)\phi_{1}(\textbf{r}),
\end{eqnarray}
where $\textbf{r}$ is the coordinate vector on $\partial\Gamma$, and
$\Omega(\textbf{r})$ is the solid angle subtended by $\partial\Gamma$ at $\textbf{r}$.
For the surface of spheres and faces of cuboids,
$\Omega(\textbf{r})=2\pi$, while for the edges of cuboids, $\Omega(\textbf{r})=\pi$, and for
the corners of cuboids, $\Omega(\textbf{r})=\pi/2$.
With $\phi_{2}$ known on $\partial\Gamma$, its known values on $\partial\Gamma$ are then applied as a Dirichlet boundary condition
for the Laplace equation below for $\phi_{2}$ in $\Gamma$:
\begin{equation} \label{laplace}
\nabla^{2}\phi_{2}=0.
\end{equation}

The weak form for Eq.~(\ref{poisson}) and its associated boundary condition, Eq.~(\ref{bc_poisson}) is
\begin{equation} \label{poisson_weak_form}
\int_{\Gamma}\nabla\psi\cdot\nabla\phi_{1}\,d^3r=\int_{\Gamma}\nabla\psi\cdot\textbf{M}({\textbf r})\,d^3r,
\end{equation}
where the test function $\psi\in H^1(\Gamma)$, and $H^1(\Gamma)$ is the usual Sobolev space.
The weak form of the Laplace equation, Eq.~(\ref{laplace}), is simply
\begin{equation}
\int_{\Gamma}\nabla\psi\cdot\nabla\phi_{2} \,d^3r=0.
\end{equation}

\subsection{Electrostatic problem}
A large class of systems in materials science involve polarizable metal oxide colloidal or nanoparticles in a dielectric continuum. The presence of a particle exerts an electrostatic field on nearby particles, leading to their polarization, which then propagates throughout the entirety of the system in a cooperative manner. Under some circumstances, polarization interactions can in fact lead to attractive forces between charged particles having the same charge, a feature that has been revealed in striking detail for metalic oxide micro-particles\cite{jaeger2015}.
Here we consider polarizable dielectric bodies $p_i$ embedded in an infinite dielectric medium $\Omega$.
Dielectric polarization arises on the surface of the bodies when the relative permittivities of the bodies are different from that of the medium.
Bound surface charges are then induced because
of the discontinuity of the electrostatic field across the interfaces between bodies and medium.
In the following, we first review key equations that facilitate implementation of the BEM.
For details on the derivation of the method, we refer to Refs.~\onlinecite{allen2001,luijten2014jcp}.
The relative permittivity of the medium is $\epsilon_{m}$, the relative permittivity of $i$th body is $\epsilon_{i}$,
and the point charge carried by the $i$-th particle is $Q_i$. In this method, $Q_i$ is represented by an equivalent surface free charge density $\sigma_{f}$ on the surface of each body (see section IV.I in Ref.~\onlinecite{luijten2014jcp}).
In the particular case of spherical bodies, $\sigma_{f}=Q_i/(4\pi r_i^2)$ on the $i$th body with radius $r_i$.
The induced bound surface charge density $\sigma_{b}$ can then be calculated by solving
\begin{equation} \label{eq:linear_system}
A\sigma_{b}=b,
\end{equation}
where
\begin{equation} \label{Ax}
A\sigma_{b}\equiv\bar{\epsilon}_{i}\sigma_{b}+\epsilon_{0}\Delta\epsilon_{i}\textbf{E}_{b}\cdot\textbf{n},
\end{equation}
with
\begin{equation} \label{eq:Eb}
\textbf{E}_{b}=\frac{1}{4\pi\epsilon_{0}}\int_{\partial\Gamma}\frac{\textbf{r}-\textbf{r}'}{{|\textbf{r}-\textbf{r}'|}^3}\sigma_{b}(\textbf{r}')\,d^2\textbf{r}'.
\end{equation}
In the above equations, $\bar{\epsilon}_{i}=(\epsilon_{i}+\epsilon_{m})/2$, $\Delta\epsilon_{i}=\epsilon_{m}-\epsilon_{i}$,
$\epsilon_{0}$ is the permittivity of vacuum,
$\textbf{n}$ is the outward-pointing normal unit vector at the interface between a body and the medium,
and $\textbf{E}_{b}$ is the electrostatic field that arises from the bound surface charge density $\sigma_{b}$.
The integral is performed only on the surfaces of all dielectric bodies ($\partial\Gamma$).
The right-hand side $b$ in Eq.~(\ref{eq:linear_system}) is obtained from
\begin{equation} \label{rhs}
b=(1-\bar{\epsilon}_{i})\sigma_{f}-\epsilon_{0}\Delta\epsilon_{i}\textbf{E}_{f}\cdot\textbf{n},
\end{equation}
with
\begin{eqnarray} \label{eq:Ef}
\textbf{E}_{f}=\frac{1}{4\pi\epsilon_{0}}\Big(\int_{\partial\Gamma}\frac{\textbf{r}-\textbf{r}'}{{|\textbf{r}-\textbf{r}'|}^3}\sigma_{f}(\textbf{r}')\,d^2\textbf{r}'\nonumber\\
+\int_{\Omega}\frac{\textbf{r}-\textbf{r}'}{{|\textbf{r}-\textbf{r}'|}^3}\frac{\rho_{f}(\textbf{r}')}{\epsilon_{m}}\,d^3\textbf{r}'\Big),
\end{eqnarray}
where $\textbf{E}_{f}$ is the electrostatic field that arises only from the free charges, $\sigma_{f}$ is the surface free charge density on the surface of the dielectric bodies,
and $\rho_{f}$ is the volume free charge density in the medium $\Omega$.
In our examples, only free charges carried by the dielectric bodies are considered, and there are no free volume charges in the medium,
so the second term in Eq.~(\ref{eq:Ef}) is absent.
Once we obtain the total surface charge density $\sigma=\sigma_f+\sigma_b$ and total electrostatic field $\textbf{E}=\textbf{E}_f+\textbf{E}_b$ on the surface of the dielectric bodies ($\partial\Gamma$),
the force vector acting on the $i$th body, $\textbf{F}_{i}$, can be calculated by a boundary integral over the surface of $i$-th body,
  \begin{equation}
\textbf{F}_{i}=\int_{\partial p_{i}}\textbf{f}(\textbf{r})\,d^2\textbf{r}
\end{equation}
where $\textbf{f}(\textbf{r})$ is the surface force density on the surface of the body,
\begin{equation}
    \textbf{f}(\textbf{r})=\epsilon_{m}\sigma \textbf{E}(\textbf{r}).
\end{equation}

\section{Implementations}

In the magnetostatic problem, a standard FEM is used to solve the Poisson and Laplace equations.
The magnetic bodies are first discretized and meshed, in our case using the CUBIT Geometry and Mesh Generator\cite{blacker1994cubit}; the mesh is then read by libMesh.
Shape functions and weights for Gaussian quadrature points are provided by libMesh as well.
For the Laplace equation, a Dirichlet boundary condition is
enforced using a penalty method\cite{babuvska1973finite}. The system of equations is then solved
using GMRES, as provided by libMesh and its underlying PETSc\cite{petsc-user-ref} library.
For the boundary integral, the strategy is to create a separate boundary mesh on the fly during the computations. The boundary mesh is extracted from the boundary of the volume mesh of all bodies using libMesh's functionalities,
and we then transfer nodal values of the boundary source term from the boundary of the volume mesh to the corresponding DoFs
on the boundary mesh before executing the boundary integration.

In the electrostatic problem, because dielectric polarization occurs at the interface,
only the surface of dielectric bodies, rather than their volume, is discretized with a triangular mesh using CUBIT.
Constant monomial shape functions is used to approximate variables such as surface charges and electrostatic fields in each element.
The boundary integration on each element is done using a one-point quadrature rule:
the quadrature point in each element is set to be the centroid of that element,
and the weight is the surface area corresponding to that element.
If the dielectric bodies are spheres, the surface area is calculated analytically as the spherical area;
if the dielectric bodies are not spheres, the surface area is approximated by the flat area of each triangular element.
The point where variables are approximated in each element is set to be the same as the quadrature point.
The linear system of equations Eq.~(\ref{eq:linear_system}) is solved using GMRES in a matrix-free form provided by libMesh and PETSc, i.e.,
we do not explicitly form the matrix $A$, rather, we define the action of matrix-vector multiplication $A\sigma_b$ that is used in each GMRES iteration.
The most computationally expensive part in the matrix-vector multiplication is calculating the electrostatic fields, which is a boundary integral with the kernel multiplied by the surface charge densities [see Eqs.~(\ref{eq:Eb}) and (\ref{eq:Ef})].
In a direct method, the computational complexity of the matrix-vector multiplication is $O(N^2)$.
In Ref.~\onlinecite{luijten2014jcp,luijten2015jcp}, a fast Ewald solver is used to reduce the complexity to $O(NlogN)$;
in our implementation, FMM is used to further reduce the complexity to $O(N)$.
In the following subsections, boundary integration and parallelization are presented.

\subsection{Integrals accelerated by FMM}

\subsubsection{Magnetostatic problem}
Here, the relevant integral is given in Eq.~(\ref{fk_integral}). 
The boundary mesh
is constructed for and applied only to the boundary integration, and it also enables us to couple the three solvers for Eqs.~(\ref{poisson})--(\ref{laplace}) in a single simulation.
We transfer nodal values of $\phi_{1}$ from the boundary of the volume mesh to the corresponding DoFs
on the boundary mesh before executing the boundary integration. After $\phi_{2}$ is evaluated on the boundary mesh,
we transfer its nodal values from the boundary mesh to the boundary of the volume mesh before imposing the Dirichelet
boundary condition. Note that the boundary mesh itself is 3D, but it uses
elements with 2D linear shape functions. We evaluate $\phi_{2}$ on nodal points of the boundary mesh using a five-point Gauss quadrature rule
and the resulting discretized form of Eq.~(\ref{fk_integral}) is then
\begin{eqnarray} \label{fk_integral_discrete}
\phi_{2}(\textbf{r}_{i})=\frac{1}{4\pi}\sum_{j=1}^{N}\frac{(\textbf{r}_{i}-\textbf{r}_{j})\textbf{n}(\textbf{r}_{j})}{{|\textbf{r}_{i}-\textbf{r}_{j}|}^3}\phi_{1}(\textbf{r}_{j})w(\textbf{r}_{j})\nonumber\\
+\left(\frac{\Omega(\textbf{r}_{i})}{4\pi}-1\right)\phi_{1}(\textbf{r}_{i}),
\end{eqnarray}
where $i$ is the Dof index on the boundary mesh,
$j$ is an index for the quadrature points, and $N$ is the total number of quadrature
points in all boundary elements. Note that $N$ depends on the number of boundary elements and the order
of the Gauss quadrature rule. Finally, $w(\textbf{r}_{j})$ is the Gauss quadrature weight at each quadrature
point. We can see that the integral at the $i$-th Dof on the boundary mesh is discretized as a
summation over all quadrature points, which is the first term in Eq.~(\ref{fk_integral_discrete}),
with values of $\phi_{1}$ and Gauss quadrature weights $w(\textbf{r}_{i})$ at quadrature points multiplied by the kernel
\begin{equation} \label{kernel}
K(\textbf{r}_i,\textbf{r}_j)=\frac{(\textbf{r}_i-\textbf{r}_j)\cdot\textbf{n}(\textbf{r}_j)}{{|\textbf{r}_i-\textbf{r}_j|}^3}.
\end{equation}
We also denote the artificial physical value at $j$-th quadrature point as
\begin{equation}
\sigma_j=\phi_{1}(\textbf{r}_{j})w(\textbf{r}_{j}).
\end{equation}
In the implementation of the summation in Eq.~(\ref{fk_integral_discrete}) using a kernel-independent FMM in ScalFMM, the normal unit vector at the quadrature points cannot be directly applied in the interpolation of the kernel.
There are two methods to include the normal unit vector. In the first, the kernel in Eq.~(\ref{kernel}) is split into three components,
\begin{equation}
K(\textbf{r}_i,\textbf{r}_j)=\frac{(x_i-x_j)n_{x,j}}{{|\textbf{r}_i-\textbf{r}_j|}^3} + \frac{(y_i-y_j)n_{y,j}}{{|\textbf{r}_i-\textbf{r}_j|}^3} + \frac{(z_i-z_j)n_{z,j}}{{|\textbf{r}_i-\textbf{r}_j|}^3},
\end{equation}
and we rewrite the summation as
\begin{widetext}
\begin{equation} \label{sum_split}
\begin{split}
f(\textbf{r}_i)&=\sum_{j=1}^N\frac{x_i-x_j}{{|\textbf{r}_i-\textbf{r}_j|}^3}\sigma_j n_{x,j} + \sum_{j=1}^N\frac{y_i-y_j}{{|\textbf{r}_i-\textbf{r}_j|}^3}\sigma_j n_{y,j}
+ \sum_{j=1}^N\frac{z_i-z_j}{{|\textbf{r}_i-\textbf{r}_j|}^3}\sigma_j n_{z,j}\\
&=\sum_{j=1}^{N}K_1(\textbf{r}_i,\textbf{r}_j)\sigma_{1,j} + \sum_{j=1}^{N}K_2(\textbf{r}_i,\textbf{r}_j)\sigma_{2,j}
+\sum_{j=1}^{N}K_3(\textbf{r}_i,\textbf{r}_j)\sigma_{3,j}.
\end{split}
\end{equation}
\end{widetext}
In this method, the summation is then carried out using three kernels, and
the components of the normal unit vector are incorporated in three artificial
physical values associated with each kernel, as shown in Eq.~(\ref{sum_split}). The second method incorporates the normal unit vector during the particle-to-multipole (P2M)  stage in ScalFMM. According to the method of kernel-independent FMM and Eq.~(6) in Ref.~\onlinecite{Fong2009}, the summation in Eq.~(\ref{sum_fmm}) can be expressed as
\begin{equation} \label{p2m}
f(x_i)\approx\sum_{l=1}^n S_n(\bar{x}_l,x_i)\sum_{m=1}^n K_r(\bar{x}_l,\bar{y}_m)\sum_{j=1}^N \sigma_j\frac{\partial S_n(\bar{y}_m,y_j)}{\partial y_j},
\end{equation}
where $n$ is the order of Chebyshev interpolation, $S_n(\bar{x}_l,x_i)$ is the interpolating function for node $\bar{x}_l$, and $K_r$ is the $1/r$ kernel such that the kernel in Eq.~(\ref{kernel}) can be written as
\begin{equation}
K(\textbf{r}_i,\textbf{r}_j)=\frac{\partial K_r(\textbf{r}_i,\textbf{r}_j)}{\partial \textbf{n}(\textbf{r}_j)}.
\end{equation}
The idea of the second method is to use the usual $1/r$ kernel, and to apply the normal unit vector as a normal derivative to the interpolating function (the last term in Eq.~\ref{p2m}), which can be implemented in the P2M stage of FMM.
One advantage with this approach is that ScalFMM is highly optimized for the symmetric $1/r$ kernel, which further reduces the computational cost of the boundary integration.
However, this second method requires modifications to the current ScalFMM source code, which are beyond the scope of our work. Therefore, we currently incorporate the normal unit vector using only the first method.

In our implementation of the boundary integral using ScalFMM, we used the target source model (TSM): the nodal points are treated as target points, and quadrature points in every element
are treated as source points, so that target points are different from source points. These target and source points are then inserted into the octree\cite{octree} with information about
their coordinates, particle type (target/source), and the artificial physical value [$\sigma_{1,j}$,$\sigma_{2,j}$,$\sigma_{3,j}$ in Eq.~(\ref{sum_split})] associated with source points.
We avoid the singularity of the integral
by setting a tolerance radius around each target point: if the distance between target and
source point is less than a small number, {\em e.g.,} $10^{-5}$, then its contribution to the integral is set to zero.
During the FMM computations, the height of the octree, which represents the balance between near-field and far-field calculations,
is tuned to achieve optimal performance of the FMM calculation. In general, the more particles in the octree, the larger its height is.

\subsubsection{Electrostatic problem}
For the numerical implementation of the boundary integrals that arise when calculating electrostatic fields, we use Eq.~(\ref{eq:Eb}) as an example for the discretization.
The electrostatic field vector at the $j$th element that arises from the bound surface charge density is
\begin{equation} \label{pol_discret}
    \textbf{E}_{b}^{j}=\frac{1}{4\pi\epsilon_{0}}\left(\sum_{k=1, k\neq j}^{N}\frac{\textbf{r}_j-\textbf{r}_k}{{|\textbf{r}_{j}-\textbf{r}_{k}|}^3}\sigma_{b}^{k}a_{k}+\sqrt{\frac{a_{j}}{\pi}}H_{j}\textbf{n}_j\right),
\end{equation}
where $N$ is the total number of boundary elements (here also the Dof because of the constant monomial shape function used here),
$a_{j}$ and $a_k$ are the surface areas of the $j$-th and $k$-th element,
$H_{j}$ is the mean curvature at $j$-th element, and
$\textbf{n}_{j}$ is the unit outward-pointing normal vector at the centroid of the $j$-th element.
The first term in Eq.~(\ref{pol_discret}) is the contribution from all source elements except for the target element itself,
and the second term is a correction term that replaces the original singular term when $j=k$. This approximate correction works reasonably well for arbitrary geometries (see section IV.B in Ref.~\onlinecite{luijten2014jcp}).
For spherical surfaces, $H_j=1/r$, where $r$ is the radius of sphere to which the $j$-th element belongs; \textcolor{red}{for cylindrical lateral surfaces, $H_j=1/(2r)$, where $r$ is the radius of the cylinder; for flat surfaces, $H_j=0$, which means the correction term is zero.}

The kernel used in the summation in Eq.~(\ref{pol_discret}) is
\begin{equation} \label{kernel_pol}
K(\textbf{r}_i,\textbf{r}_j)=\frac{\textbf{r}_i-\textbf{r}_j}{{|\textbf{r}_i-\textbf{r}_j|}^3}.
\end{equation}
Note that this kernel is different from that in Eq.~(\ref{kernel}) for the magnetostatic problem, because it contains no component of the normal unit vector.
It is also evident from Eqs.(\ref{Ax}) and (\ref{rhs}) that the normal unit vector is associated with target points rather than source points, and is outside of the boundary integral.
We also note that the above kernel in Eq.~(\ref{kernel_pol}) is the spatial derivative of $1/r$ kernel.
Because the $1/r$ kernel and its spatial derivatives, i.e. the forces from the $1/r$ potential, are already implemented in ScalFMM,
we can take advantage of its highly optimized algorithm specifically designed for the symmetric $1/r$ kernel to perform boundary integrals for the electrostatic problem, which is more efficient than our current implementation for the boundary integrals in the magnetostatic problem.

\subsection{Parallelization}

In our parallel implementation, we use a serial mesh, i.e. every processor has a copy of the boundary mesh (and volume mesh in the magnetostatic problem), but the data structures, such as matrices and vectors, used to solve the linear system of equations are partitioned and distributed among all processors automatically in libMesh during the parallel solution
using PETSc. A parallel mesh is needed only if storing mesh data on a each processor consumes too much memory, which may be the case for problems with billions of DoFs. We use METIS\cite{metis} as the partitioner for data structures associated with the mesh.

The magnetostatic problem has a volume mesh for solving the Poisson and Laplace equations [Eqs.~(\ref{poisson}) and (\ref{laplace})] and a boundary mesh
for the boundary integral [Eq.~(\ref{fk_integral})]. If we then were to use the same
partitioner for the volume mesh and boundary mesh, poor load balancing would result. That is because even if the volume mesh is almost uniformly
partitioned, the boundary mesh is generally not.
Therefore, we assign one partitioner to the volume mesh and a separate partitioner to the boundary mesh,
which significantly improves the parallel performance of the boundary integration, especially when the number
of processors is large.
For the electrostatic problem, only one partitioner is necessary and the main parallelization is implemented in building the right-hand side $b$,
performing matrix-vector multiplication in each GMRES iteration, and calculating the force vectors on the dielectric bodies, i.e.,
every processor calculates its own contribution to the right-hand side, to the matrix-vector product, and to the force vectors.

For the parallelization of FMM in the magnetostatic problem, one strategy is to partition only the source points, i.e., every processor has a
copy of the coordinates of all target points on the boundary mesh, as well as coordinates, physical values of $\phi_1$, and
normal unit vectors of only local source points.
We then perform serial FMM calculations on every processor, which means that we first calculate
the contributions from local source points to all target points. We then let all processors communicate and sum up contributions from
every processor and assign integrated values to all target points.
An alternative strategy would share a similar concept, but partition only target points and keeps on every processor a local copy of information from all source points.
The specific choice between the above two strategies that is better suited for the problem at hand depends on whether the number of target points or number of source points is largest. If the number of source \textcolor{red}{(target)} points is larger than that of target \textcolor{red}{(source)} points, we
choose to partition the source \textcolor{red}{(target)} points, \textcolor{red}{\textit{i.e.,} the larger set of points,} because this can reduce \textcolor{red}{the computational time and} the memory cost of storing particles on every processor.
Most of the times we found the number of source points is larger than the number of target points, because the number of quadrature points is mostly larger than that of nodal points, so we choose to
partition the source points in our parallel implementation.

The strategy for parallelization of FMM in the electrostatic problem is similar to that in the magnetostatic problem.
The only difference comes from the fact that we use different shape functions in this problem than those employed for the magnetostatic problem. Since a constant monomial shape function is used here,
the quadrature point (source point) in each element is also the point (target point) where variables, such as charge densities and electrostatic fields, are approximated, i.e. target points are the same as source points.
In order to use the TSM model in ScalFMM and the strategy of partitioning only target/source points, all quadrature points are inserted into the octree first as target points, and then they are inserted into the same octree again as source points, along with information of associated physical values and quadrature weights.
We then partition only the target points; each processor keeps a copy of all source points along with the information of physical values and quadrature weights, and we perform FMM in serial on every processor.

We also note that ScalFMM has its own strategy for parallelization of FMM,
in which both target and source points are partitioned based on their Morton ordering\cite{bbthesis}. The idea is to map 3D points in the octree to those in a 1D array based on the z-value of each point, and then to partition all points based on the index of their 1D representation.
After partitioning, local points are inserted into the octree on each processor, and each processor then performs FMM and communicates with other processors\cite{parallel_scalfmm}. This strategy consumes less memory than
the previous two methods, but its implementation is more involved. Therefore, we have not considered it here, and will examine it in future developments of our computational approach, where we will also compare its parallel efficiency with methods that partition only target/source points.
In the following section, we test our computational approaches for the magnetostatic and electrostatic problems by applying them to several model problems, and we compare the results with analytical solutions. These comparisons serve to demonstrate the accuracy, efficiency, and scalability of our parallel approach.

\section{Results and discussion}

\subsection{Magnetostatic problem}

\begin{figure}[h]
\includegraphics[width=0.45\textwidth]{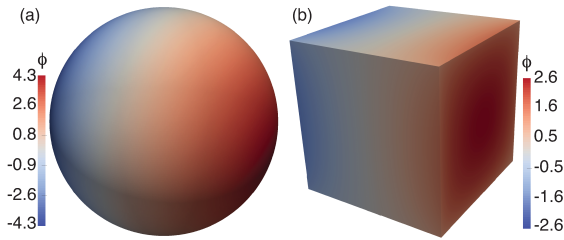}
\centering
\caption{Model magnetostatic problems for magnetic bodies with homogeneous unit magnetization. The color contour shows the magnetic potential ($\phi$) distribution in the bodies, (a) the unit sphere, and (b) the unit cube.}
\label{fig2_fk_color}
\end{figure}

Two model magnetostatic problems are considered here to verify our proposed methods and their implementation: a uniformly magnetized
unit radius spherical body, and a uniformly magnetized unit cubical body, with unit length side (see Fig.~\ref{fig2_fk_color}). For demonstration purposes, the magnetization density within each body is set to be $\textbf{M}=(1,0,0)$.
After solving the magnetostatic problem using our computational approach, the distribution of the magnetic potential within each body is obtained, as shown in Fig.~\ref{fig2_fk_color}.
For the unit sphere, the magnetic potential varies linearly along the direction of the magnetization density, and the stray field (demagnetizing field) is calculated by
\begin{equation}
\textbf{H}=-\nabla \phi,
\end{equation}
which is the (negative) gradient of the magnetic potential, and it is uniform everywhere inside the sphere in this particular case.
The analytical solution of the stray field along the direction of magnetization density is $4\pi/3$ in Gaussian units, see Ref.~\onlinecite{fk1990}.
For the unit cube, the magnetic potential and stray field are nonuniform within the body. Instead, we compute the stray-field energy and compare our result with analytical solutions. The stray field energy is calculated as
\begin{equation}
E_{s}=-\frac{1}{2}\int_{\Gamma}\textbf{M}\cdot\textbf{H}\,d^3\textbf{r}.
\end{equation}
The analytical value of the stray field energy in a unit cube with unit homogeneous magnetization density is $2\pi/3$ [$\mu_{0} M_s^2$] (see Refs.~\onlinecite{Abert2013,aharoni1998}), with $M_s$ the saturation magnetization density.
We calculate the relative error for the stray field energy, for example, as
\begin{equation} \label{relative_error}
\varepsilon_{rel}=\left|\frac{E_{s,numeric}-E_{s,analytic}}{E_{s,analytic}}\right|
\end{equation}
\begin{figure}[h]
\includegraphics[width=0.5\textwidth]{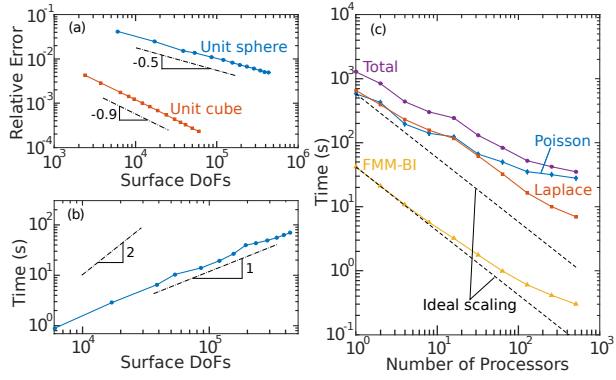}
\centering
\caption{(a) \textcolor{red}{Relative errors} of the stray field in the unit sphere, and the stray field energy in the unit cube calculated from simulations compared to their analytical values, as a function of surface DoFs of the magnetic bodies. The inset numbers indicate convergence rates. (b) Computational time spent on the boundary integral as a function of
surface DoFs, tested using a single CPU core on Intel Xeon E5-1607 v3 @ 3.1GHz. The inset numbers 1 and 2 indicate the scaling of $O(N)$ and $O(N^2)$ respectively. (c) Strong scaling of our computational approach for modeling a unit sphere with 7.4M volume DoFs and 70K surface DoFs, tested on high performance computer Blues at Argonne National Laboratory (ANL).}
\label{fig1_fk}
\end{figure}

\begin{figure}[h]
\includegraphics[width=0.38\textwidth]{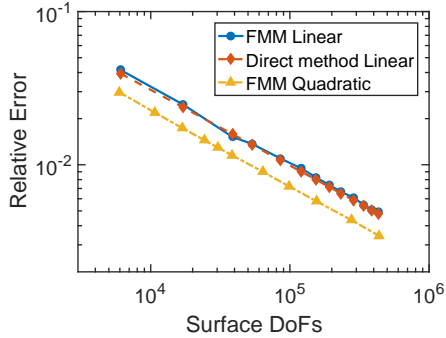}
\centering
\caption{\textcolor{red}{Relative errors of the stray-field energy in the unit sphere as a function of surface DoFs of the magnetic body using FMM and a direct method with linear shape functions, and FMM with quadratic shape functions.}}
\label{fig_fk_error}
\end{figure}

Figure~\ref{fig1_fk}(a) shows the relative errors as a function of surface DoFs; they decay linearly in the log-log plot.
For a certain number of surface DoFs, the relative error for the unit cube is smaller than that for the unit sphere; also the convergence rate for the unit cube is larger than that for the unit sphere from \textit{a posteriori} estimates.
The larger errors associated with the sphere calculations are due to the larger discretization error that arises when using flat elements and linear shape functions to describe a curved geometry. They do not arise because of selected settings in ScalFMM; indeed, Fig.~\ref{fig_fk_error} shows that the difference between the relative errors by FMM and and the direct method is small, and that using quadratic shape functions reduces the discretization errors.
One way to decrease the relative errors in modeling curved surfaces is to adaptively refine the surface mesh so that surface DoFs and computational cost increase simultaneously; another way is to use curved elements or higher-order methods\cite{curved_element_2013} to approximate variables on curved surfaces.

\begin{figure}[h]
\includegraphics[width=0.53\textwidth]{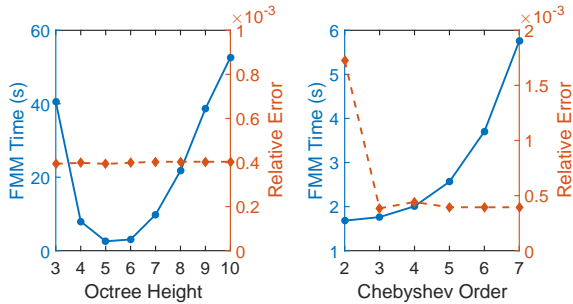}
\centering
\caption{\textcolor{red}{FMM computational time and relative errors of the stray-field energy for the unit cube as a function of octree height (left panel) and the Chebyshev order (right panel). The volume and surface DoFs of the unit cube considered here are 373248 and 30248 respectively.}}
\label{fig_fmm_tree_order}
\end{figure}

Figure~\ref{fig1_fk}(b) shows the time spent on computing the boundary integral by FMM as a function of surface DoFs.
The order of Chebyshev polynomial is 5 in all cases.
The time increases linearly as the surface DoFs increase, and $O(N)$ scaling is confirmed.
Note that the total number of points inserted in the octree is the sum of surface DoFs
(target points) and the number of quadrature points (source points).
We use a fifth-order quadrature rule for the boundary integration, so the number of source points in the octree is roughly 8 times that of target points.
In using FMM to accelerate computation of boundary integrals, it is critical to choose the appropriate octree height to achieve optimal FMM performance, because it controls the balance between the near-field and far-field computation in FMM.
The optimal octree height depends on the total number of points in the octree and the size of the cubic box, and they are determined by trial runs.
As DoFs increases, the optimal octree height generally increases. In Fig.~\ref{fig1_fk}(b), two small humps are observed when the surface DoFs is 53K and 192K, and they indicate changes in the octree height.
The trade-off between FMM settings (the octree height and the order of Chebyshev polynomial) and performance is shown in Fig.~\ref{fig_fmm_tree_order}. The relative errors of the stray-field energy change little when octree height changes, and the accuracy in FMM is mainly determined by the Chebyshev order. Increasing the Chebyshev order improves accuracy, but the computational time increases. We also refer to Ref.~\onlinecite{Fong2009} for more details on the trade-offs between FMM settings and performance.

We also tested the parallel efficiency of a simulation of magnetized spherical bodies with
7.4M volume DoFs and 70K surface DoFs. Figure~\ref{fig1_fk}(c) shows the strong scaling
for time spent on the Poisson solver, the boundary integral using FMM, the Laplace solver, and
the total time for solving the full problem. For the Poisson solver and Laplace solver,
the time includes that spent on construction of the block-Jacobian preconditioner as well as
the time spent on solving the linear system of equations using GMRES. We tested up to 512 processors and the time spent on each
part decreases as the number of processors decreases. When the number of processor is larger
than 64, we observe that the decay rate of time on the Poisson solver is slower than
that on the Laplace solver and the boundary integral, while the decay rates of time on the
Laplace solver and boundary integral are very similar. Algebraic multigrid preconditioners\cite{petsc-user-ref}, such as gamg and hypre, for the Poisson solver were also tested with the result that total computational time and strong scaling are similar to those using block-Jacobian pre-conditioner.
Constructing pre-conditioners using gamg or hypre takes longer time than using a block-Jacobian method, but the number of GMRES iteration is much less using gamg and hypre than using block-Jacobian.
For static problems, the total times are similar using different pre-conditioners but, for study of time-dependent problems, if the global matrix remains the same between different time steps, then pre-conditioning using gamg or hypre is recommended.

\subsection{Electrostatic problem}

\begin{figure}[h]
\includegraphics[width=0.35\textwidth]{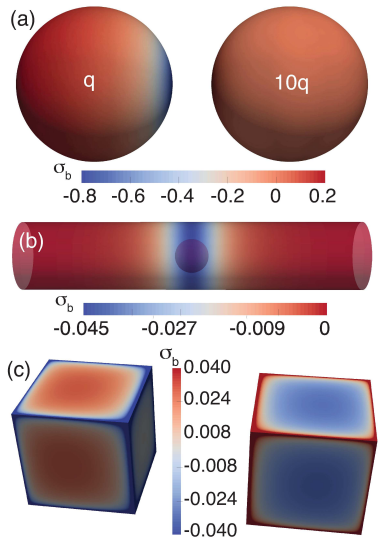}
\centering
\caption{Model electrostatic problems for (a) two polarizable spherical dielectric bodies, (b) a polarizable dielectric sphere inside a cylindircal cavity in a dielectric medium, and (c) two polarizable dielectric cubes. The color maps show the induced {\em bound} surface charge density, ($\sigma_b$), in units of $q/(\epsilon_0R^2)$. In (a), the two spheres have the same relative permittivity $\epsilon_{1}=\epsilon_{2}=15$ and the same radius $r_1=r_2=R$, the relative permittivity of the medium is $\epsilon_m=1$, the point charges are $Q_1=10 q$ (body on the right), $Q_2=q$ (body on the left), and the center to center distance is $d/R=2.5$. In (b), the relative permittivities of the sphere and the cylindrical cavity are unity, while the relative permittivity of the medium is 2. The point charge on the sphere is $q$ while the cylinder has no surface free charge density, the length and radius of cylinder is $10R$ and $R$; the radius of the sphere is $0.5R$ and its located at the center of the cylinder's axis. In (c), the side of cubes is $R$, the center to center distance is $d/R=2.0$, the surface free charge densities are $-1/4\pi$ for the left cube and $1/4\pi$ for the right cube, and the relative permittivities of cubes and medium are the same as those in (a).}
\label{fig_color_dielectric}
\end{figure}

\begin{figure}[h]
\includegraphics[width=0.5\textwidth]{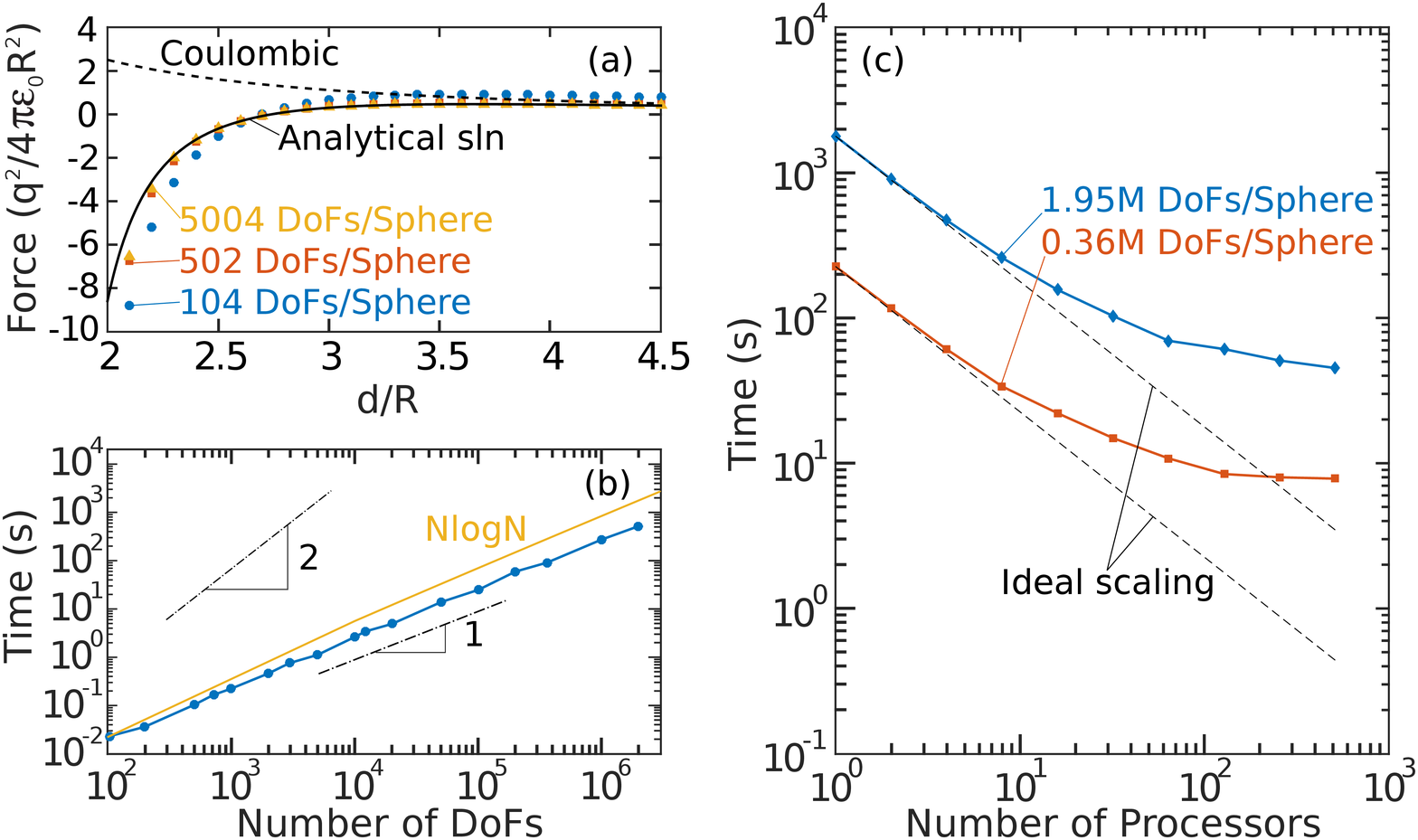}
\centering
\caption{(a) Comparison between numerical and analytical results of calculating the electrostatic forces
between two dielectric spherical bodies as a function of their center to center distance. Negative value means attractive force.\cite{sedimentation} (b) The scaling behaviors and computational time spent on solving for the surface bound charge densities and
calculating body forces on particles using our numerical approach as a function of surface DoFs, tested on a single CPU core of Intel Xeon E5-1607 v3 @ 3.1GHz. The inset numbers 1 and 2 indicate the scaling of $O(N)$ and $O(N^2)$ respectively. For comparison purpose, it is assumed that the break-even point between $O(N)$ and $O(NlogN)$ is at 100 DoFs.
(c) Strong scaling of solving the model electrostatic problem using our numerical approach tested on Blues at ANL.}
\label{fig2_polarization}
\end{figure}

We verified our method and implementation with a few model calculations for which there are analytical as well as numerical literature solutions: specifically, two dielectric spheres with surface-bound free charges, and a sphere inserted in a cylindrical cavity in a dielectric medium. In addition, we tested our method on a model problem for which there exists no analytical solution in closed form, namely two dielectric cubes with uniform surface free charges. The results are shown in Fig.~\ref{fig_color_dielectric}, and Fig.~\ref{fig_color_dielectric}(a) shows the solution for the problem of two spherical dielectric bodies.
We also varied the center-to-center distance between two spheres and calculated the electrostatic forces at each distance using our numerical approach and an analytical expression\cite{Qin2016,jian_theory} (see Fig.~\ref{fig2_polarization}(a)). Figure~\ref{fig_color_dielectric}(b) shows the solution for the dielectric sphere inside a cylindrical cavity in a dielectric medium. The polarization effect on the sphere is turned off by setting its relative permittivity to be the same as that of the medium; note that this problem is equivalent to that of a point charge inside a dielectric cylinder. Our results are in good agreement with analytical solutions as well as with  previous numerical results from Ref.~\onlinecite{allen2001}. Figure~\ref{fig_color_dielectric}(c) shows results for the two dielectric cubes. In particular, the bound surface charge densities accumulate on the corners and edges of the cubes, rather than on the faces with a max/min value of +/- 0.49. To better show the polarization effect on cubic faces, the color range is set to +/- 0.04 in Fig.~\ref{fig_color_dielectric}(c).

\begin{figure}[h]
\includegraphics[width=0.42\textwidth]{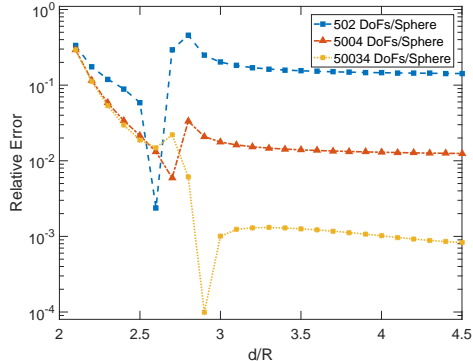}
\centering
\caption{\textcolor{red}{Relative errors of the body forces on unit dielectric spheres calculated from simulations compared to their analytical values for different DoFs/Sphere, as a function of center to center distance between two spheres.}}
\label{fig_pol_error}
\end{figure}

Figure~\ref{fig2_polarization}(a) shows the comparison between our numerical method and an analytical
method of calculating body forces on two dielectric spherical bodies as a function of their center-to-center
distance (particle separation). For the numerical method, the surface DoFs on each body are varied to test the accuracy, while the relative tolerance for the iterative solver GMRES is set to $10^{-5}$ in all cases \textcolor{red}{(further decreasing the relative tolerance improves numerical results little)}.
\textcolor{red}{The relative errors for the body force based on Eq.~\ref{relative_error} are shown in Fig.~\ref{fig_pol_error}. As the number of degrees of freedom increases, the relative errors decreases significantly when $d/R\geq2.8$. They decrease marginally when $d/R\leq2.5$, and there are known difficulties in calculating forces accurately when two spheres are very close to each other because of large discretization errors, and adaptive mesh refinement is a technique to efficiently improve the accuracy in this situation\cite{luijten2014jcp}. The relative errors do not always decrease when $2.5<d/R <2.8$; under the same number of degrees of freedoms, for example 5004 DoFs/Sphere, they are even larger when $2.7<d/R<3.0$ than those when $d/R\geq3.0$. This is because the body forces are nearly zero at these locations as shown by analytical solutions, and they are orders of magnitude smaller than those at other locations and thus more sensitive to discretization errors, which makes the relative errors large while the absolute error is very small.}

Including polarization effects results in very different electrostatic interactions compared to those when only bare Coulombic interactions are included, especially when particle separations are small. These issues have been reported in past theoretical and numerical studies. In the particular case considered in Fig.~\ref{fig2_polarization}, even though the surface free-charge densities on both bodies are all positive, the electrostatic interactions may exhibit attractive
forces between the bodies when their center-to-center distance ($d/R$) is less than 2.74. Note that the bare Coulomb
interaction consists of only repulsive forces, so it is crucial that the effects of dielectric polarization be included
in calculations of electrostatic interactions for simulations of colloidal bodies, lest the underlying physics not be captured correctly.

Figure~\ref{fig2_polarization}(b) shows the total computational time spent on assembling the right-hand-side of Eq.~(\ref{eq:linear_system}), solving the linear system of equations iteratively using GMRES, and calculating the body forces on one particle.
As before, the order of Chebyshev polynomial is 5 in all cases.
We can see that as surface DoFs increase, the computational time increases linearly, which is similar to that of the boundary integral using FMM in the magnetostatic simulation, and it confirms the
$O(N)$ complexity of our method. In each case, because the distance between two particles is different from those in other cases, the box size for the octree is also different. The octree height in each case is tuned to achieve optical FMM performance, and the height generally increases as the surface DoFs increase.
\textcolor{red}{Compared to PPPM and the particle-mesh Ewald method\cite{PME} (PME) with $O(NlogN)$ complexity, FMM-based methods exhibit better computational complexity; Figure~\ref{fig2_polarization}(b) shows that they have a clear advantage in high-performance computing of large-scale simulations with millions of unknowns. In addition, FMM-based methods are mesh-free approaches and more efficient for heterogeneous systems (\textit{e.g.,} systems with electrostatic polarization), where the mesh-based approaches such as PPPM and PME that require an evenly spaced FFT-mesh become slow and memory-intensive\cite{pre-fmm-fft}.}

Figure~\ref{fig2_polarization}(c) shows strong scaling for up to 512 processors for solving the electrostatic problem
when the surface DoFs on each particle are 0.36~M and 1.95~M. Overall, the computational times decrease as number of processor increases, and they are slightly larger than for
ideal scaling when the number of processors is less than four. The difference between the computational time and ideal scaling increases as the number
of processors increases. For the case with 0.36~M DoFs using 256 and 512 processors, the computational times were found to reach a plateau, while the strong scaling shows better performance for 1.95~M DoFs.
In order to investigate what prevented good scaling ({\em i.e.,} closer to ideal) for larger numbers of processors, we performed a detailed profiling of our code. We found that the time spent on inserting all source particles and FMM calculations on all source points dominate the total time, while the communication time spent on gathering all information of source points to every processor is much smaller.
This observation shows that the plateau in strong scaling is the result of our parallel strategy of partitioning only target points and keeping a copy of all source points on every processor, so that all source points need to be inserted into the octree. This dominates the insertion time.
In addition, FMM calculations have an $O(N)$ complexity, where $N=N_t/np+N_s$ since we only partition target points, $N_t/np$ is the number of target points per processor and $N_s$ is the number of all source points.
In the electrostatic case, $N_t=N_s$ because of the one-point quadrature rule used in the boundary integration.
So the FMM calculation time is dominated by calculations over source points, even if there are much fewer target points when the number of processors ($np$) is reasonably large.
One strategy to recover better scaling for larger numbers of processors would be to create a separate partitioner for FMM, and to distribute both target and source points to every processor according to their Morton index. There will then be two partitioners in the simulation: one for FMM, and one for boundary elements. The challenge is then to create a mapping between the two partitioners to couple FMM with the boundary element calculation. Because our current parallel implementation shows very good scaling up to 128 processors, and quite good scaling for 1.95~M DoFs up to 512 processors, we believe our implementation is more than adequate for most applications. Further improvements to our parallel implementation will be performed in future work.

\section{Conclusions}

We have developed accurate, efficient, and scalable parallel computational approaches for integral 
problems based on the open source libraries libMesh and ScalFMM. The problems considered here all contain integrals of Green's functions
over the boundary of the relevant domains, which exhibit a computational complexity of $O(N^2)$ in direct
implementations. To facilitate wider usage of these methods, we accelerated the computation of
the boundary integrals using a kernel-independent FMM to achieve $O(N)$
computational complexity, and also with a memory cost of $O(N)$, resulting in significant gains in computational speed as well as in efficient memory utilization.

We note that FMMs have been used previously for boundary integrals and BEM\cite{fmm_bem_1995,fmm_bem_2006,Yokota2011}, as well as combining with GMRES \cite{gmres_fmm_1998,gmres_fmm_2006}. However, there are few publicly available parallel codes using these techniques, in particular
highly efficient and scalable ones. 
Our work is aimed at providing such a code framework, which we developed with the following features in mind:
first, the code must run in parallel for large-scale problems with good parallel performances;
\textcolor{red}{second, it must be available for general distribution, at http://ime-code.uchicago.edu as part of the Continuum-Particle Simulation Suite (COPSS) from the Midwest Integrated Center for Computational Materials (MICCoM);} third, it should provide convenient interfaces for computational researchers who are not familiar with acceleration techniques or parallelizations to focus on the underlying physical problems, as opposed to software engineering.
Our computational approaches open up numerous possibilities and will enable efficient modeling for a variety of large-scale 
many-body problems, such as dynamics of colloids and interactions of magnetic bodies, to explore new physics and functions of materials.
Further improvements can be made on our computational approaches, such as using the parallelization strategy in ScalFMM with the concept of Morton ordering and compare its performance with
partitioning only target/source points as in our current implementation. Furthermore, for the boundary integral in the magnetostatic problem,
the normal derivative in the Green's function could be implemented in the P2M stage in ScalFMM, which will take advantage of the highly-optimized symmetric $1/r$ kernel in ScalFMM to further accelerate the boundary integrations.

\section*{Acknowledgements}
X.J. acknowledges support by U. S. DOE, Office of Science under Contract No. DE-AC02-06CH11357. The work by J.H.-O., J.J.d.P., and O.H. was supported by MICCoM, as part of the Computational Materials Sciences Program funded by the U.S. Department of Energy, Office of Science, Basic Energy Sciences, Materials Sciences and Engineering Division. We gratefully acknowledge the computing resources provided on Blues and
Fusion, high-performance computing clusters operated by the Laboratory Computing Resource Center at Argonne National Laboratory.

\bibliography{mybibfile}

\begin{thebibliography}{48}
\bibitem [{oct()}]{octree}%
  \BibitemOpen
  \href@noop {} {\bibinfo  {journal} {3D space is partitioned into a tree
  structure in which each node has eight children}\ }\BibitemShut {NoStop}%
\end{thebibliography}

\end{document}